\documentclass[prb,aps,twocolumn,showpacs]{revtex4}
\usepackage{epsfig}
\usepackage{graphicx}
\usepackage{amsmath}

\begin{document}

\title[{\it Ab-initio} study of pressure-induced changes in liquid Si]
{Pressure-induced structural and dynamical changes in 
liquid Si. An {\it ab initio} study}

\author{A. Delisle$^{1}$, D.J. Gonz\'alez$^{2}$ and M.J. Stott$^{1}$}

\address{$^{1}$ Department of Physics, Queen's University, 
Kingston, Ontario, CANADA}

\address{$^{2}$Departamento de F\'\i sica Te\'orica, Universidad 
de Valladolid, Valladolid, SPAIN}

\date{\today}

\begin{abstract}
The static and dynamic properties of liquid Si at high-pressure 
have been studied using the orbital free {\em ab-initio} molecular 
dynamics method. Four thermodynamic states at pressures of 4, 8, 14 
and 24 GPa are considered, for which X-ray scattering data are 
available. The calculated static structure shows qualitative 
agreement with the available experimental data. We analize the remarkable 
structural changes occurring between 8 and 14 GPa along with its reflection 
into several dynamic properties. 
\end{abstract}


\maketitle

\section{Introduction.}

The intriguing properties of Silicon along with its technological 
importance have stimulated intensive theoretical 
\cite{StillingerWeber,Tersoff,JankHafner,Wang,Virkkunen,
Stich1,Stich2,Cheli,DGS1} and experimental 
\cite{Gabathuler,Waseda1,Waseda2,Takeda,Hosokawa1,Hosokawa2,
FunamoriTsuji} work. Its high-density forms include the 
semiconducting and covalent crystalline and amorphous phases 
and the metallic liquid phase. Upon melting it undergoes a 
semiconductor-metal transition, a density increase of $\approx$ 
10\% and significant changes in the local structure which evolves 
from an open one, with a tetrahedral fourfold coordination, to a 
liquid structure with $\approx$ sixfold coordination. 
In crystalline Si (c-Si)
the semiconducting diamond structure contracts with pressure and 
transforms at 12 GPa \cite{Jamieson} to the metallic white-tin 
structure  and then to the metallic simple hexagonal structure at 
16 GPa \cite{Spain}. The local structure of liquid Si (l-Si) at the
triple point (TP) is somewhat similar to high-pressure forms of 
c-Si, and it has been suggested that l-Si might 
consist of a mixture of diamond-type and white-tin-type structures 
with the proportion of the latter increasing with pressure. 

Within this backdrop, Funamori and Tsuji \cite{FunamoriTsuji} have 
recently carried out 
X-ray (XR) diffraction experiments to determine the static structure of 
l-Si at pressures of 4, 8, 14 and 23 GPa and temperatures about 50 K 
above the melting point at the pressure. From an analysis of the static 
structure factors $S(q)$ and the associated pair distribution functions 
$g(r)$, Funamori and Tsuji \cite{FunamoriTsuji} concluded that 
l-Si up to 8 GPa has a local structure intermediate between the 
diamond-type and the white-tin-type. But, between 8 and 14 GPa drastic 
structural changes were noted with l-Si transforming to a denser 
structure similar to that of l-Sn at ambient pressure  
as evinced by the strong similarities between the 
$S(q)$'s of l-Sn and l-Si at 14 GPa.

Prompted by these experimental developments, we have performed  
an {\it ab-initio} molecular dynamics (AIMD) study of several 
static and dynamic properties of compressed l-Si at the thermodynamic 
states addressed by Funamori and Tsuji \cite{FunamoriTsuji}. Of 
particular interest is the reflection of the reported structural 
changes in the dynamic properties. Our AIMD method is based on density 
functional theory \cite{HK-KS} (DFT) which, for given nuclear positions, 
allows the calculation of the ground state electronic energy and 
yields the forces on the nuclei via the Hellmann-Feynman theorem, 
enabling MD simulations in which the nuclear positions evolve 
according to classical mechanics whereas the electronic subsystem 
follows adiabatically. Most AIMD methods are based on the Kohn-Sham 
(KS) form of DFT (KS-AIMD methods) which treats the electron kinetic 
energy exactly, but which at present, poses heavy computational 
demands limiting the size of the systems to be studied as well as 
the simulation times. Some of these constraints can be relaxed by 
the so-called orbital-free {\it ab-initio} molecular dynamics 
(OF-AIMD) method, which approximates the electron kinetic energy
but disposes of the electronic orbitals of the KS formulation. The
method allows simulations in which the number of variables describing 
the electronic state is greatly reduced so that larger samples 
(several hundreds of particles) can be studied for longer 
simulation times (tens of ps).

Theoretical studies of l-Si have mainly focused on static structural
properties for thermodynamic states near the TP. Most studies were 
classical MD simulations using effective interatomic potentials 
constructed either empirically by fitting to experimental data 
\cite{StillingerWeber,Tersoff} or derived from some approximate 
theoretical model. \cite{JankHafner,Wang,Virkkunen}. Recently,   
KS-AIMD calculations \cite{Stich1,Stich2,Cheli} have been 
reported which address electronic and static properties. Stich 
{\it et al} \cite{Stich1} and Chelikowsky {\it et al} \cite{Cheli} have 
reported 
KS-AIMD calculations for l-Si for 64 particles, using non-local 
pseudopotentials and the local density approximation.  
A subsequent calculation \cite{Stich2} used 350 particles and an 
improved treatment of electron exchange and correlation. 
Recently, we have carried out an OF-AIMD simulation \cite{DGS1} 
for l-Si near the TP for 2000 particles using a first principles local 
pseudopotential. Both static and dynamic properties were calculated 
with results in good agreement with the available experimental data, 
supporting the validity of the OF-AIMD for treating systems such as 
l-Si which show some remnants of covalent bonding and are not fully 
metallic. 

On the experimental side, besides the aforementioned XR experiments of 
Funamori and Tsuji \cite{FunamoriTsuji}, we also quote the availability 
of both neutron scattering (NS) \cite{Gabathuler} and 
X-ray (XR) \cite{Waseda1,Waseda2,Takeda} diffraction data as well as 
the recent inelastic X-ray scattering (IXS) data of 
Hosokawa {\it et al} \cite{Hosokawa1,Hosokawa2} which    
have provided information on the dynamic structure of l-Si near TP. 

In the next section the orbital-free {\em ab-initio} molecular 
dynamics (OF-AIMD) scheme is described briefly 
with emphasis on the electronic kinetic energy functional and the 
local pseudopotential used to characterize the electron-ion interaction. 
In section \ref{results} the results of the {\em ab initio} simulations 
for several static and dynamic properties are presented and compared with 
the available experimental data. Finally, conclusions are drawn and 
ideas for further improvements are suggested.

\section{Theory.}
\label{theory}

The OF-AIMD method used in this study is described fully in earlier work
\cite{GGLS}, and has previously been used to study 
l-Si near the TP \cite{DGS1}. 
In summary, an explicit density functional for the electronic 
energy is minimized iteratively for each ion configuration, the forces 
on the ions are found using the Hellman-Feynman theorem and the ion 
positions and velocities are updated by solving Newton's equations. 
The approximate electron kinetic energy functional which correctly 
gives the Thomas-Fermi and linear response limits is based on the von 
Weizs\"{a}cker term plus a correction which uses an averaged density 
\cite{GGLS}. The local electron-ion pseudopotential was 
constructed, for each thermodynamic state, according to the procedure 
described in reference \cite{GGLS}.

Simulations have been performed for l-Si in the four thermodynamic 
states listed in Table \ref{states}. These correspond to pressures of 
4, 8, 14 and 23 GPa and temperatures about 50 K above the melting point
for each pressure \cite{FunamoriTsuji}. Each simulation used 2000 
ions in a cubic cell with periodic boundary conditions and size 
appropriate for the ionic number density, $\rho_i$. The square root of 
the electron density was expanded in plane waves up to a cutoff energy
$E_{\rm Cut}=15.75$ Ryd. The Verlet leapfrog algorithm with a timestep of 
$3.5\times 10^{-3}$ ps was used to update the ion positions and velocities.
Equilibration lasted 10 ps. and the calculation of properties was made 
averaging over a further 65 ps. For comparison, we mention that the 
KS-AIMD simulations for l-Si near the TP lasted 1.2 ps. \cite{Stich1}
0.9 ps. \cite{Stich2} and 1.0 ps. \cite{Cheli}, which precludes its 
application to the study of most dynamical properties. 

Several liquid static properties were evaluated during the simulation:
pair distribution function, static structure factor and bond angle 
distribution, as well as various dynamic properties, both single-particle 
ones: velocity autocorrelation function, mean square displacement, and 
collective ones: intermediate scattering functions, dynamic structure 
factors, longitudinal and transverse currents. The calculation of the 
time correlation functions (CF) was performed by taking time origins 
every five time steps. Several CF are also dependent on the wave vector
q $\equiv$$\mid {\bf q} \mid$.

\begin{table}[h]
\caption{\label{states} Input data for the different thermodynamic states 
studied in this work. $\rho_i$ is the total ionic number density 
and T is the temperature which have been 
taken from Ref. \cite{FunamoriTsuji}. } 
\begin{tabular}{lllll}
& $P (GPa) $ & $\rho_i$ (\AA$^{-3}$) & T ($^0$K) \\ 
\hline
& 4 & 0.058 & 1503 &  \\
& 8 & 0.060 & 1253 &  \\ 
& 14 & 0.067 & 1093 &  \\ 
& 23 & 0.071 & 1270 &  \\ 
\hline
\end{tabular}
\end{table}

\section{Results}
\label{results}

\subsection{Static properties.}

\begin{figure}
\begin{center}
\mbox{\psfig{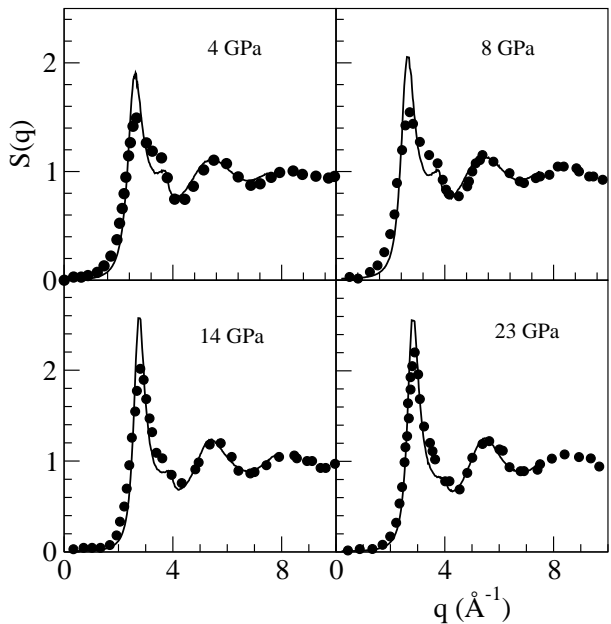}}
\end{center}
\caption{Static structure factor of l-Si at different 
high pressures. Full circles: experimental X-ray diffraction 
data \cite{FunamoriTsuji}. 
Continuous line: OF-AIMD simulations.} 
\label{sqfig4}
\end{figure}

The simulations yield directly the pair distribution 
function, $g(r)$, and the static structure factor $S(q)$. 
Figure \ref{sqfig4} shows the calculated $S(q)$'s 
along with the corresponding XR data of Funamori 
and Tsuji \cite{FunamoriTsuji}. 
The experimental $S(q)$'s show changes with increased pressure. The 
main peak grows in intensity and its position ($q_p$) increases 
monotonically, whereas the position of the second peak 
decreases between 8 and 14 GPa; the distinctive shoulder at 
the high-q side of the main peak, shrinks smoothly and practically 
vanishes at 23 GPa. These changes are also reflected in $g(r)$. The 
position of the main peak, ($r_p$), identified with the average 
nearest neighbor distance, decreases with pressure except for
an increase between 8 and 14 GPa; the position of the second peak 
decreases monotonically with pressure. 

These features are displayed 
qualitatively in the calculated $S(q)$'s and $g(r)$'s although 
there are some quantitative discrepancies with the experimental data. 
Figure \ref{sqfig4} shows that the OF-AIMD $S(q)$'s overestimate 
the intensity of the main peak and slightly underestimate the 
shoulder. Otherwise the positions of the peaks as well as the amplitudes 
of the subsequent oscillations are accounted for fairly well. A more 
detailed comparison with experiment is provided in Table \ref{results1} 
which summarizes most of this structural information. This agreement with 
experiment is similar to that achieved in earlier orbital-free simulations 
\cite{DGS1} and in KS-AIMD \cite{Stich1,Stich2,Cheli} calculations 
performed for for l-Si near TP.

Based on their experimental data for $S(q)$ and $g(r)$, Funamori and Tsuji 
\cite{FunamoriTsuji} have argued that l-Si undergoes a high pressure 
structural transformation between 8 and 14 GPa. Whereas l-Si contracts 
with pressure up to at least 8 GPa by reducing the bond length, as 
measured by $r_p$, the increase in $r_p$ between 8 and 14 GPa suggests a 
structural change with an increase in the coordination number (CN). An
estimate of the CN may be obtained  by integrating the radial distribution 
function (RDF), $4\pi r^2 \rho_i g(r)$, up to the position of the first
minimum, $r_m$, in the RDF \cite{Cusak,McGreevy}. The results from the 
calculated RDF in Table \ref{results1} show that CN grows with compression 
but with an abrupt increase from 8 to 14 GPa. Funamori and Tsuji 
\cite{FunamoriTsuji} obtained CN values also given in Table 
\ref{results1} by integrating their experimental RDF up to 3.1 \AA $\;$ 
for all the states, and a similar growth between 8 and 14 GPa is seen. 
Had we used the same $r_m$  = 3.1 \AA $\;$ as Funamori and Tsuji did for 
calculating the CN, then the agreement with the "experimental" values 
would have been even better but, taking $r_m$ as the position of the 
first minimum is thought to be more soundly based.

Values for the isothermal compresibility, $\kappa_T$, have been obtained
from  $S(0) = \rho_i k_B T \kappa_T$ by using a least squares fit to 
calculated the $S(q)$ for $q$-values up to 0.8 \AA$^{-1}$ and 
extrapolating to $q \to 0$. Results are given in Table \ref{results2}. 
Although no experimental results are available, the OF-AIMD 
calculation for l-Si near the TP yielded $\kappa_T$ = 3.0, which is 
rather close to the experimental 
value \cite{BaidovGitis} of 2.8 (in $10^{-11}$ m$^2$ Nw$^{-1}$ units).  

\begin{figure}
\begin{center}
\mbox{\psfig{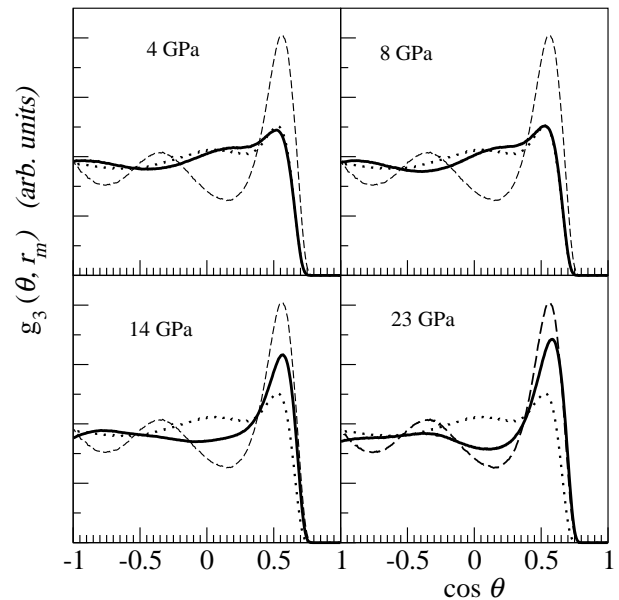}}
\end{center}
\caption{Bond-angle distribution function, $g_3(\theta, r_m)$,  
of l-Si at different high pressures (full line). The dotted    
and dashed lines stand respectively for l-Si and l-Al at their 
respective TP. } 
\label{bondangle4}
\end{figure}

Further structural information is provided by higher order correlation 
functions such as the bond-angle distribution function, $g_3(\theta,r_m)$, 
where $\theta$ is the angle between two vectors joining a reference 
particle with two neighboring particles at a distance less than $r_m$.  
In a simple liquid metal such as Al, the $g_3(\theta,r_m)$, has peaks at 
around $\theta$ $\approx$ $60^o$ and $120^o$, which are close to those 
expected for a local icosahedral arrangement \cite{Balubook} ($\theta$ 
$\approx$  $63.5^o$ and $116.5^o$). In contrast, for l-Si near the TP  
both the OF-AIMD \cite{DGS1} and KS-AIMD \cite{Stich1,Stich2,Cheli} 
calculations for $g_3(\theta,r_m)$ have yielded two maxima centered 
around $\theta \approx 60^o$ and $89^o$. This double-peak feature has been 
interpreted as a manifestation of tetrahedral bonding and higher 
coordinated atoms both contributing to the first coordination shell.  
In illustration, figure \ref{bondangle4} shows the OF-AIMD results for 
the $g_3(\theta,r_m)$ of l-Si and l-Al near their TP's. \cite{DGS1,GGLS}.  
OF-AIMD results for compressed l-Si are also included in the figure, and
a gradual evolution with pressure towards the simple liquid metal 
distribution is seen. There is little change up to 8 GPa although the 
wide maximum at $\theta \approx$ $89^o$ has moved to slightly smaller 
$\theta$ values which may indicate less tetrahedral bonding. 
But, as pressure increases from 8 to 14 GPa, there is a qualitative 
change in the $g_3(\theta,r_m)$ whose shape moves closer to that of the 
simple liquid metals, and at 23 GPa the positions of the maxima for l-Si
and l-Al are rather similar. 

\medskip

\begin{table}[h]
\caption{\label{results1}
Calculated values of $q_p$ (\AA$^{-1}$), $r_p$ (\AA), $r_m$ (\AA)
and coordination number (CN),
for the different states. The numbers in parenthesis are the corresponding
experimental data from Ref. \cite{FunamoriTsuji} }
\begin{tabular}{llllllllll}
& $P (GPa)$ $\; \;$ & $q_p$ & & $r_p$ & & $r_m$ &  CN & 
CN \tablenotemark[1] &   \\
\hline
& 4 &  2.61 (2.67) & & 2.52 (2.46) & & 3.03 &  6.6 (6.8) & 6.9 &  \\
& 8 &  2.65 (2.72) & & 2.50 (2.42) & & 3.07 &  7.2 (7.1) & 7.3 &  \\
& 14 & 2.78 (2.82) & & 2.51 (2.46) & & 3.20 &  9.6 (8.5) & 8.9 &  \\
& 23 & 2.84 (2.88) & & 2.47 (2.43) & & 3.28 &  11.0 (9.2) & 9.5 &  \\
\hline
\end{tabular}
\tablenotetext[1]{Calculated by integrating the RDF up to $r_m$ = 3.1 \AA }
\end{table}

\medskip

\subsection{Dynamic properties.}

\subsubsection{Collective dynamics.}

The intermediate scattering function, $F(q, t)$, contains both spatial and 
temporal information on the collective dynamics of density fluctuations. 
It is defined as

\begin{equation}
F(q, t) = \frac{1}{N} \left \langle 
\left( \sum_{m=1}^N  
e^{-i {\vec q}{\vec R}_m(t + t_0)} \right)
 \; 
\left( \sum_{l=1}^N e^{i {\vec q}{\vec R}_l(t_0)} \right) \right \rangle
\end{equation}

\noindent 
Its frequency spectrum is the dynamic structure 
factor, $S(q, \omega)$, which has experimental relevance due 
to its connection with the scattered intensity in inelastic X-ray or 
neutron experiments. The calculated $F(q, t)$ for compressed l-Si exhibit 
an oscillatory behaviour up to $ q \approx (3/5) \; q_p$, with the 
amplitude diminishing for larger $q$-values. This oscillatory behaviour 
is typical of simple liquid metals found by either computer simulation 
\cite{GGLS,TorBalVer,Shimojo2,Kambayashi} or from theoretical models 
\cite{Litio} and gives rise to a well defined inelastic peak in 
$S(q, \omega)$.

\begin{figure}
\begin{center}
\mbox{\psfig{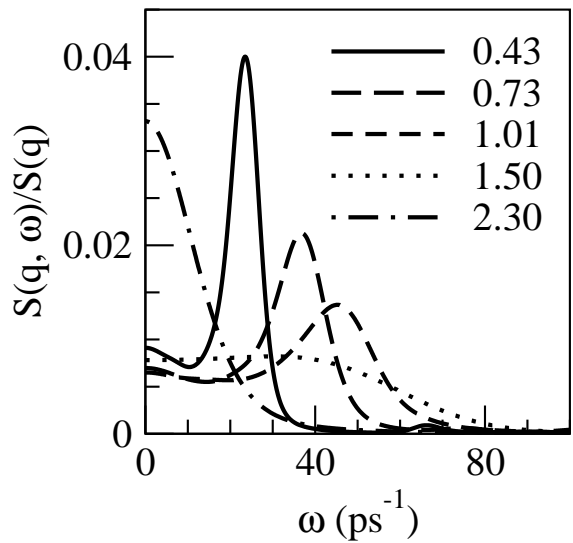}}
\end{center}
\caption{Dynamic structure factor $S(q, \omega)$ 
at several $q$-values (in \AA$^{-1}$ units), for l-Si 
at 8 GPa.  }
\label{Sqw}
\end{figure}

The $S(q, \omega)$, obtained by a time FT of $F(q, t)$, exhibit for all 
the states,  well defined sidepeaks which are indicative of collective 
density excitations. This is illustrated in figure \ref{Sqw}, which
shows calculated $S(q, \omega)$ for l-Si at 8 GPa for several $q$ values. 
The general shape of $S(q, \omega)$ is qualitatively similar at 
equivalent $q/q_p$ values for all the compressed states, and no specific 
feature of $S(q, \omega)$ has been identified whose variation would mark 
the structural transformation occurring somewhere between 8 and 14 GPa. 

\begin{figure}
\begin{center}
\mbox{\psfig{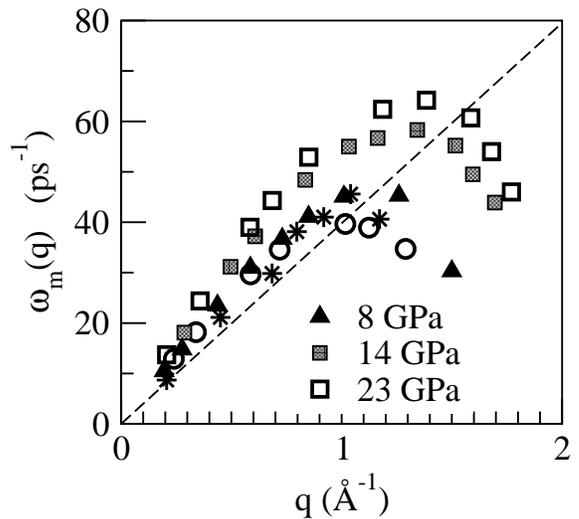}}
\end{center}
\caption{Dispersion relation for the peak positions, $\omega_m(q)$, from 
the calculated $S(q, \omega)$, for l-Si at  
different high pressures. 
The figure also includes the 
calculated (open circles) (Ref. \cite{DGS1})  
and experimental (asterisks) (Ref. \cite{Hosokawa1}) 
results for l-Si near the triple point (T=1740 K). 
Dashed line: linear dispersion with the hydrodynamic sound 
velocity, v=3977 m/s., at the triple point. }
\label{disper4}
\end{figure}

For all the states the sidepeaks in $S(q, \omega)$ persists up to $q 
\approx (3/5) q_p$, which is a feature shared by both l-Si \cite{DGS1} and 
the simple liquid metals near their TP's. \cite{GGLS,Balubook} 
The dispersion relations, $\omega_m(q)$, of the density fluctuations 
have been obtained from the positions of these sidepeaks. They are plotted 
in figure \ref{disper4} for the 8, 14 and 23 GPa states together with 
the calculated OF-AIMD \cite{DGS1} results and experimental data 
\cite{Hosokawa1} for l-Si near its TP. The curves look qualitatively 
similar but there is a marked difference between $\omega_m(q)$ for 8 GPa
and less, and those for 14 and 23 GPa. In addition, the slope of the 
dispersion gives a $q$-dependent adiabatic sound velocity, $c_s(q)$,  
which in the limit $q \to 0$ reduces to the bulk adiabatic sound velocity, 
$c_s$. This has been estimated by fitting a straightline to the low-$q$ 
region of the $\omega_m(q)$'s  and the results are given in Table 
\ref{results2}. The $c_s$ increase with pressure but a steeper rise from 
8 to 14 GPa will be seen.

\bigskip

\begin{figure}
\begin{center}
\mbox{\psfig{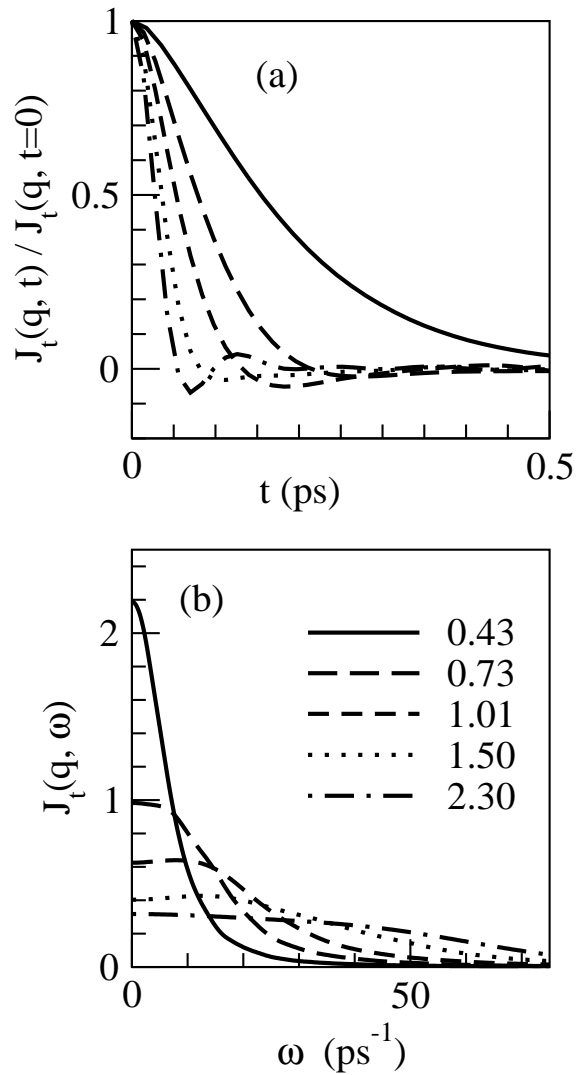}}
\end{center}
\caption{(a) Transverse current correlation function $J_t$(q,t) 
at several $q$-values (in \AA$^{-1}$ units), for l-Si  at 8 GPa.  
(b) Same for $J_t$(q,$\omega$). } 
\label{Ctqtw1253}
\end{figure}

\begin{figure}
\begin{center}
\mbox{\psfig{file=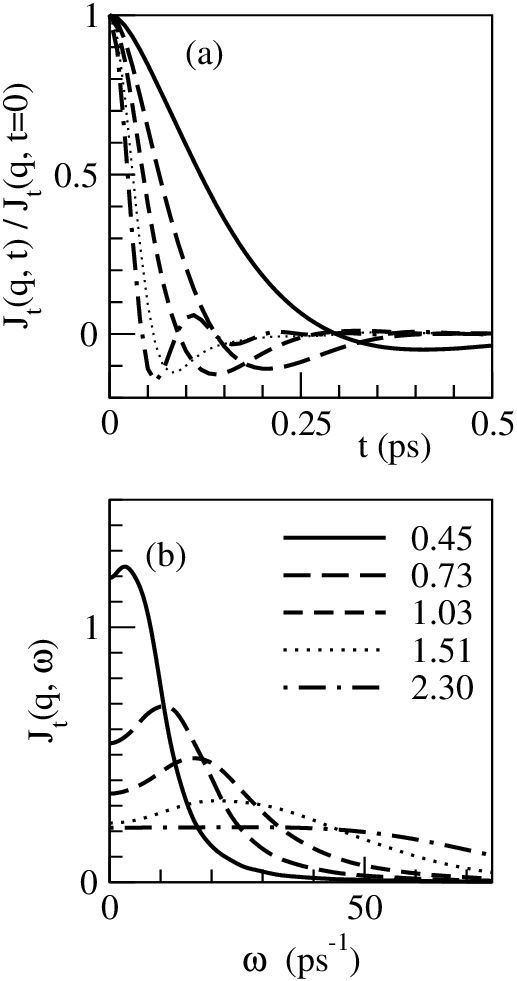,angle=0,width=75mm,clip}}
\end{center}
\caption{(a) Transverse current correlation function $J_t$(q,t) 
at several $q$-values (in \AA$^{-1}$ units), for l-Si  
at 14 GPa.  
(b) Same fo $J_t$(q,$\omega$). } 
\label{Ctqtw1093}
\end{figure}

The transverse current correlation function, $J_t(q,t)$, provides 
information on shear modes and is not directly related to any measurable 
quantity. It can only be obtained from either theoretical models or 
computer simulations, but it is known that its shape evolves from a 
gaussian, in both $q$ and $t$, at the free particle ($q \to \infty$) 
limit, towards a gaussian in $q$ and  an exponential in $t$ in the 
hydrodynamic limit ($q \to 0$), i.e. 

\begin{equation}
J_t(q \to 0, t) = \frac{1}{\beta m} e^{-q^2 \eta \mid t \mid /m \rho_i} \; ,
\label{Jtqthyd}
\end{equation}

\noindent 
where $\eta$ is the shear viscosity coefficient, $\beta=(k_B T)^{-1}$ and  
$m$ is the atomic mass. In both small and large $q$ limits $J_t(q, t)$ 
is always positive, but for intermediate $q$-values there is a   
more complicated behavior with well-defined oscillations 
\cite{GGLS,Balubook,BoonYip}. 
Calculated $J_t(q, t)$ for several $q$-values are shown in figures 
\ref{Ctqtw1253}-\ref{Ctqtw1093} for l-Si at 8 and 14 GPa respectively. The 
most noteworthy effect on $J_t(q, t)$ of an increasing pressure is reflected on
the oscillations which have a smaller amplitude and last for appreciably 
shorter times at the lower presures. Its consequences  
are apparent 
in the frequency spectra, $J_t(q, \omega)$, which are plotted in the lower
panels of figures \ref{Ctqtw1253}-\ref{Ctqtw1093}. For both 14 and 23 GPa the  
$J_t(q, \omega)$ exhibits an inelastic peak which appears at low
$q$-values ($\approx$ 0.45 \AA$^{-1}$) and persists up to 
about $q$ = 2.50 \AA$^{-1}$. However for 8 GPa the 
inelastic peaks appear for a appreciably smaller range 
(0.85 \AA$^{-1}$ $\leq$ q $\leq$ 1.50 \AA$^{-1}$) while for 
4 GPa there are no inelastic peaks. 
This absence of peaks in $J_t(q, \omega)$
is also a feature of l-Si near the TP, but is at variance with the 
behaviour of a large number of different liquids such as hard sphere 
systems \cite{BoonYip}, Lennard-Jones liquids \cite{Balubook,BoonYip} 
and simple liquid metals \cite{GGLS,Balubook,Kambayashi} near 
melting, for which $J_t(q, t)$ oscillates and the associated 
$J_t(q, \omega)$, has an inelastic peak over some range of $q$-values.  
The inelastic peak in $J_t(q, \omega)$ is associated with propagating 
shear waves which seem to be absent in l-Si up to somewhere 
between 4 and 8 GPa. 
However, it must be noted that whereas in simple liquid metals near 
melting the shear waves last up to $q$ $\approx$ $3 q_p$, in the 
case of l-Si at 14 and 23 GPa they appear up to 
$q$ $\approx$ $0.9 q_p$. 

The shear viscosity coefficient, $\eta$, can be calculated from $J_t(q,t)$
using the memory function representation \cite{GGLS,Palmer,BaBroJedVa}  

\begin{equation}
\tilde{J_t}(q, z)= \frac{1}{\beta m} 
\left [ z + \frac{q^2}{\rho_i m} \;  \tilde{\eta}(q, z)\right ]^{-1} \; ,
\end{equation}

\noindent where the tilde denotes the Laplace transform, and 
$\tilde{\eta}(q, z)$ is a generalized shear viscosity coefficient. The  
$\int_0^{\infty}\,dt\,J_t(q, t)$ when normalized gives $\beta m \; 
\tilde{J_t}(q, z=0)$, from which $\tilde{\eta}(q, z=0)$ are obtained 
and extrapolated to $q=0$ to give the usual shear viscosity coefficient, 
$\eta$. Results are given in Table \ref{results3}. Although no comparison
can be made with experimental results, the calculated values are 
considered reliable because the application of this approach to l-Si 
near its TP gave $\eta$=0.75 $\pm$ 0.15 GPa $\cdot$ ps, in reasonable 
agreement with the available experimental 
data \cite{SasakiKimura}, $\eta_{exp}$=0.58-0.78  GPa $\cdot$ ps. 
It is noteworthy that once more $\eta$ undergoes 
an abrupt change as the pressure increases from 8 to 14 GPa. 

\bigskip

\medskip

\begin{table}[h]
\caption{
\label{results2}
Calculated values of S(q $\to$ 0), isothermal 
compressibility $\kappa_T$ (in $10^{-11}$ m$^2$ Nw$^{-1}$ units) 
and  adiabatic sound velocity ($c_s$) 
for the different states.  } 
\begin{tabular}{llllll}
& $P (GPa)$ $\; \;$ & S(q $\to$ 0) & $\kappa_T$ ( $10^{-11}$ $m^2$ Nw$^{-1}$) 
&  $c_s$ $(m/s)$ & \\ 
\hline
& 4 &  0.0180 $\pm$ 0.003  & 1.50 $\pm$ 0.3 & 5100  & \\ 
& 8 &  0.0135 $\pm$ 0.003  & 1.30 $\pm$ 0.3 & 5400  & \\ 
& 14 &  0.0095 $\pm$ 0.003 & 0.94 $\pm$ 0.3 & 6300  & \\ 
& 23 &  0.0085 $\pm$ 0.003  & 0.68 $\pm$ 0.3 & 6750  & \\ 
\hline

\end{tabular}
\end{table}

\medskip

\subsubsection{Single-particle dynamics.}

\begin{figure}
\begin{center}
\mbox{\psfig{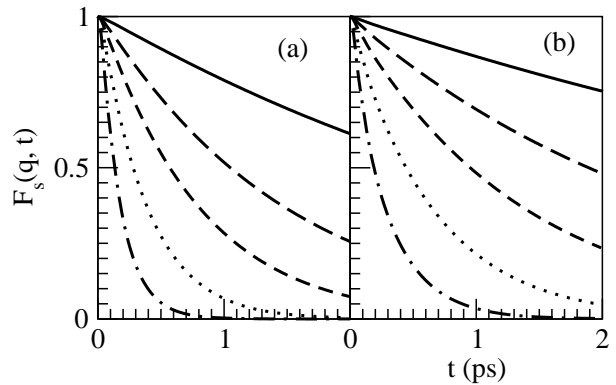}}
\end{center}
\caption{Self intermediate scattering functions, $F_s(q, t)$, at  
several $q$-values (in \AA$^{-1}$ units), for l-Si at (a) 8 GPa  and 
(b) 14 GPa. The key for the symbols is the same as in the previous figure.} 
\label{fskt4}
\end{figure}

Information about the single-particle properties is contained in the 
self-intermediate scattering function

\begin{equation}
F_s(q, t) = \frac{1}{N} \langle \sum_{j=1}^N  
e^{-i {\vec q}{\vec R}_j(t + t_0)} 
  e^{i {\vec q}{\vec R}_j(t_0)} \rangle
\end{equation}

\noindent 
and its frequency spectrum, the self-dynamic structure factor, 
$S_s(q, \omega)$, which is related to the incoherent part of the 
total intensity scattered in an INS experiment. 
The OF-AIMD results for the $F_s(q, t)$  presented in figure 
\ref{fskt4} for two states display the usual monotonic decay with 
time, and comparison of different states shows that at similar  $q/q_p$ 
values $F_s(q, t)$ decays slower with increasing pressure, with the 14 
and 23 GPa states behaving very much like the liquid simple metals 
\cite{GGLS,Balubook,Litio} near their triple points. The different rates 
of decay can be related to the differences in the self-diffusion 
coefficients. 

Closely related to the $F_s(q, t)$ is the velocity autocorrelation 
function (VACF) of a tagged ion in the fluid, $Z(t)$, which can be 
obtained as the $q \to 0$ limit of the first-order memory function of the 
$F_s(q, t)$, but more conveniently, from its definition

\begin{equation}
Z (t) = \langle \vec{v}_1(t) \vec{v}_1(0) \rangle
/ \langle v_1^2 \rangle  
\end{equation}

\begin{figure}
\begin{center}
\mbox{\psfig{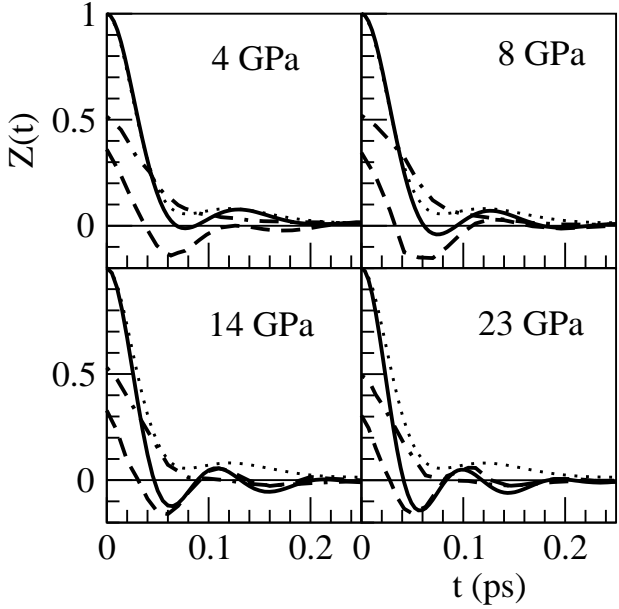}}
\end{center}
\caption{ Full line: Normalized velocity autocorrelation function $Z(t)$ for 
l-Si at different pressures. The dashed and dash-dotted lines stand for the 
respective longitudinal, $Z_l(t)$, and transverse, $Z_t(t)$, components 
as defined in equation (\ref{Ztmc}). The dotted line depicts the result 
for l-Si at triple point (T =1740 K).} 
\label{vacft4x4}
\end{figure}

\begin{figure}
\begin{center}
\mbox{\psfig{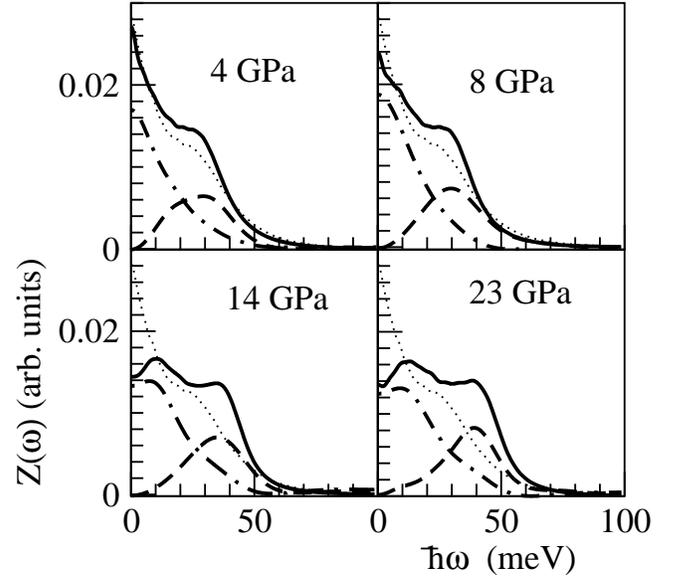}}
\end{center}
\caption{ Power spectrum, $Z(\omega)$, for 
l-Si at different pressures. The dashed and dash-dotted lines stand 
for the respective longitudinal, $Z_l(\omega)$, and 
transverse, $Z_t(\omega)$, components. The dotted line depicts the result 
for l-Si at triple point (T =1740 K).} 
\label{vacfw4x4}
\end{figure}

\noindent Figures \ref{vacft4x4}-\ref{vacfw4x4} show results for $Z(t)$ 
and for its power spectrum $Z(\omega)$. The overall shape of $Z(t)$ changes
little from the TP up to 8 GPa closely, but as the pressure is further 
increased changes occur in the range and amplitude of the oscillations 
until at 23 GPa the shape is very similar to that of the simple liquid 
metals at their TP. These changes can be explained in terms of the 
so-called "cage" effect due to backscattering from the shell of nearest 
neighbors reversing the initial velocity of a tagged ion and driving a deeper 
first minimum. This is consistent with the results for
the static structure summarized in table \ref{results1} which shows an 
open structure up to 8 GPa but a marked increase in the coordination
number at higher pressures.

The power spectra which are plotted in figure \ref{vacfw4x4}
also show significant changes between 8 and 14 GPa. $Z(\omega)$ evolves
with pressure from a shape resembling  l-Si near the TP towards a 
liquid simple metal shape with a low frequency peak and a higher frequency 
peak (or shoulder) \cite{GGLS,Balubook}. The shoulder at $\omega$ $\approx$
40 ps$^{-1}$, present at all the pressures, has been related to
vibrational remnants in the liquid of the covalent bonding \cite{Stich1}.
Below 8 GPa, low frequency diffusive modes are present. 

The self-diffusion coefficient, $D$, is readily obtained from either 
the time integral of $Z(t)$ or from the slope of the mean square 
displacement $\delta R^2(t) \equiv 
\langle | \vec{R}_1(t) - \vec{R}_1(0) |^2 \rangle$  
of a tagged ion in the fluid, as follows   

\begin{equation}
D= \frac{1}{\beta m} \int_0^{\infty} Z(t) dt\; ;\hspace{0.91 cm}
D= \lim_{t \to \infty} \delta R^2(t)/6t 
\end{equation}

\noindent
Both routes for $D$ lead to practically the same value, and the results  
are given in Table \ref{results3}. The decreasing values of $D$ with 
increasing pressure is due to the growing importance of backscattering.    
No experimental results are available for the diffusion coefficients of 
l-Si, but confidence in the results may be taken from the agreement 
between the OF-AIMD result for l-Si near the TP: $D_{\rm OF-AIMD}= 2.28 
\;$ \AA$^2$/ps, and the estimates from KS-AIMD calculations of 
Stich {\it et al} \cite{Stich1,Stich2}: $D_{\rm KS-AIMD}=2.02 \;$ 
\AA$^2$/ps. \cite{Stich1} which slightly increased to $2.4$  
\AA$^2$/ps. \cite{Stich2} when the number of particles was augmented to 
350 particles. Another KS-AIMD study by Chelikowsky {\it et al}  
\cite{Cheli} has yielded $D_{\rm KS-AIMD}= $1.90 $\;$ \AA$^2$/ps.
The results for $D$ at 4 and 8 GPa are similar to the value at the 
l-Si triple point whereas the results for 14 and 23 GPa are closer to 
those for the liquid simple metals near their triple points 
\cite{GGLS,Iida,Alemany}. This change in $D$ with pressure explains  
the different decay rates found in the $F_s(q,t)$ which decayed much 
faster at the lower pressures. Recalling the accurate gaussian 
approximation \cite{Balubook,BoonYip}, $F_s(q,t)= exp [ - q^2 \; \delta 
R^2(t) \; /6 ]$, it will be seen that a greater $D$ implies a greater  
$\delta R^2(t)$ and therefore a faster decay of $F_s(q,t)$. 

\medskip

\begin{table}[h]
\caption{Calculated values of the  
self-diffusion ($D$) and shear viscosity $\eta$ (in GPa ps) 
for the different states. } 
\label{results3}
\begin{tabular}{llllll}
& $P (GPa)$ $\; \;$  &  $D$ (\AA$^2$/ps) &  
& $\eta$ (GPa ps) &  \\ 
\hline
& 4 &  1.82 $\pm$ 0.05 & & 0.77 $\pm$ 0.10 & \\
& 8 &  1.33 $\pm$ 0.05 & & 0.84 $\pm$ 0.10  & \\
& 14 &  0.70 $\pm$ 0.03 & & 1.47 $\pm$ 0.15 & \\
& 23 & 0.70 $\pm$ 0.03 & & 1.55 $\pm$ 0.15 & \\
\hline
\end{tabular}
\end{table}

\medskip

The self-diffusion coefficient, $D$, of a macroscopic particle of 
diameter $d$ undergoing Brownian motion in a liquid of viscosity $\eta$ 
is related to $\eta$ through the Stokes-Einstein (SE) relation $\eta$ 
$D$ = $k_B T$/2$\pi$$d$. This relation has often been used on an atomic
scale to estimate $\eta$ by identifying $d$ with the position, $r_p$, of 
the main peak in $g(r)$. Using the $D$ values for 4, 8 ,14 and 23 GPa 
the relation yields $\eta$=0.72, 0.82, 1.36 and 1.59 GPa$\cdot$ps 
respectively, values rather close to the earlier OF-AIMD estimates. 
 
Gaskell and Miller \cite{GaskellMiller}, have used mode-coupling 
theory to develop a representation of the normalized VACF which has 
been used to interpret MD data in various fluids \cite{GaskellMiller,
Barrat,Baluwater}, and which sheds light on l-Si. Within this approach 

\begin{eqnarray}
\label{Ztmc}
Z (t) 
& & \approx \frac{1}{24 \pi^3} \int d{\bf q} \; f(q) 
\left[ J_l(q, t) + 2 J_t(q, t) \right] F_s(q, t) \nonumber \\ 
& & 
\equiv Z_l(t) \;  + \;  Z_t(t) \nonumber \\
\end{eqnarray}

\noindent where $J_l(q, t)$ and $J_t(q, t)$ are 
the normalized longitudinal and transverse current correlation 
functions and $f(q)$ is 

\begin{equation}
f(q)= \frac{3}{\rho_i} \frac{j_1(aq)}{aq}
\end{equation}

\noindent 
with $j_1(x)$ the spherical Bessel function of order one,  
$\rho_i$ is the ion number density and $a = ( 3/4\pi\rho_i)^{1/3}$ 
is the radius per ion. Substitution into equation (\ref{Ztmc}) of the 
OF-AIMD results for $J_l(q, t)$, $J_t(q, t)$ and $F_s(q, t)$ allows
identification of longitudinal and transverse current contributions: 
$Z_l(t)$ and $Z_t(t)$ respectively. The two contributions are plotted 
in figure \ref{vacft4x4} which shows that the oscillatory behaviour 
in the $Z(t)$ is due to $Z_l(t)$, but again the step from 8 to 14 GPa 
changes the shape of both contributions. Up to 8 GPa, $Z_t(t)$ remains 
positive for all times as a result of the positive nature of $J_t(q,t)$ 
(see figures \ref{Ctqtw1253}-\ref{Ctqtw1093}) and determines the long 
time behaviour of the $Z(t)$, however, from 14 GPa on, 
$Z_t(t)$ develops a shallow and 
broad negative minimum centered at rather long times $\approx$ 0.15 ps. 
On the other hand, $Z_l(t)$ accounts for most of 
the backscattering effect. With 
higher pressure the first minimum sharpens and moves to shorter 
times and the oscillations extend further, which are results of the 
increasing role of the "cage" effect. At 14 and 23 GPa, both components 
are similar in shape to their liquid simple metals counterparts 
\cite{GaskellMiller} with both oscillating about zero and with $Z_l(t)$ 
controlling the large $t$ behaviour of $Z(t)$. Finally, notice 
that the development of the deep minimum in the $Z(t)$ is mainly due 
to the rapid decay of $Z_t(t)$ with increasing pressure. 

The longitudinal and transverse components of the power spectrum, 
$Z(\omega)$, are shown in figure \ref{vacfw4x4}. 
The spectrum at small $\omega$ is dominated by $Z_t(\omega)$, and, 
consequently, the diffusion constant $D \propto Z(\omega =0)$ is 
completely determined by the transverse component. For 4 and 8 GPa,  
$Z_t(\omega)$ decreases monotonously but at 14 and 23 GPa the value
at zero frequency has dropped and a low-frequency peak has developed. 
Note that $Z_t(\omega)$ has no maximum for 4 and 8 GPa which are the 
states where the $J_t(q, \omega)$ shows, either no inelastic peaks (4 GPa) 
or they exist for a small range (8 GPa). 
The longitudinal component $Z_l(\omega)$ always exhibits a peak 
whose position increases slightly with increasing pressure.
This peak is responsible for the shoulder in the total $Z(\omega)$ for 
the 4 and 8 GPa states, as well as for the high-frequency peak 
for the 14 and 23 GPa states.

\begin{figure}
\begin{center}
\mbox{\psfig{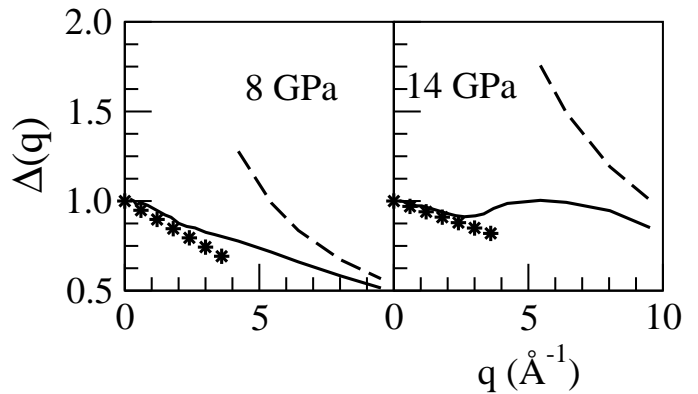}}
\end{center}
\caption{ Normalized HWHM of $S_s(q, \omega)$, relative to 
its value at the hydrodynamic limit, for l-Si at 8 and 
14 GPa. Continuous line: OF-AIMD results. 
Asterisks: Mode-coupling theory. Dashed line: free-particle limit.}
\label{Delta4}
\end{figure}

\medskip

\begin{figure}
\begin{center}
\mbox{\psfig{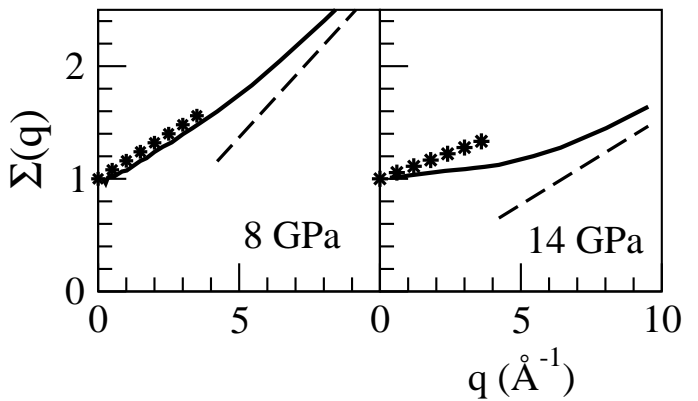}}
\end{center}
\caption{ Same as before, but for 
the normalized peak value $S_s(q, \omega=0)$, relative to 
its value at the hydrodynamic limit. }
\label{Sigma4}
\end{figure}

\medskip

By a time FT of the $F_s(q, t)$ we obtain its frequency spectrum, 
$S_s(q, \omega)$, which is known 
as the self-dynamic structure factor and is related to the incoherent 
part of the measured INS cross-section. For all $q$-values $S_s(q,\omega)$ 
decays monotonically as a function of frequency and can be 
characterized in terms of the peak value, $S_s(q, \omega = 0)$ and the
HWHM, $\omega_{1/2}(q)$. These parameters are 
frequently reported normalized with respect to the hydrodynamic 
($q$ $\to$ 0 ) limit, by introducing the dimensionless quantities 
$\Sigma(q)=\pi  q^2 D S_s(q, \omega =0)$ and 
$\Delta(q) = \omega_{1/2}(q)/ q^2 D$, where  $\omega_{1/2}(q)/ q^2$ can 
be interpreted as an effective $q$-dependent diffusion coefficient 
$D(q)$. For a simple liquid near its TP, $\Delta(q)$ usually 
oscillates whereas in a dense gas it decreases monotonically 
from unity at $q$ = 0 to the $1/q$ behaviour at large $q$ 
\cite{Balubook,TorBalVer,BoonYip}. 
Figures \ref{Delta4}-\ref{Sigma4} depict   
the OF-AIMD results for $\Delta(q)$ and $\Sigma(q)$ for 
8 and 14 GPa, as this is the pressure range where both magnitudes undergo a 
substantive change. The obtained $\Delta(q)$ 
for 14 GPa shows an oscillatory shape 
with a minimum located at $q$ $\approx$ $q_p$ which can be traced 
back to structural features ("cage" effect) which somewhat hinder the 
motion of the ions and becomes more effective at 
$q$ $\approx$ $q_p$ where the wavelength is comparable to the size of the 
cage. Conversely, that for 8 GPa resembles the situation 
of a dense gas where the "cage" effect becomes negligible leading to 
a net reduction of the diffusion coefficient \cite{Balubook,Mont} and its  
associated $\Sigma(q)$, stands very close to that of the dense gas. 

An additional check on the reliability of these results may be provided  
by the MC theory \cite{Sjogren,Sjogren2} which has already shown its 
capability to describe to experimental data for  
$\Delta(q)$ and $\Sigma(q)$ in simple liquid 
metals \cite{Mont,Cabrillo} at $q$ $\leq$ $q_p$. 
Specifically, the MC theory avers that at low-$q$ values 

\begin{eqnarray}
\label{modecp}
\Delta(q) = 1 + H(\delta) q/q^* \\ 
\Sigma(q)= 1 + G(\delta^{-1})q/q^*     \nonumber
\end{eqnarray}

\noindent where $q^* = 16 \pi m \rho_i \beta D^2$, 
$\delta= D/(D+\eta/m \rho_i)$ and $H(\delta)$ and 
$G(\delta^{-1})$ are given in Reference \cite{Mont}.

The first term in equations (\ref{modecp}) stands for the hydrodynamic 
result whereas 
the second one accounts for the coupling of mass diffusion and 
collective modes. 
Calculated values of $D$ and $\eta$ 
have been used with equations \ref{modecp} to obtain the points in 
figures \ref{Delta4}-\ref{Sigma4}. 
For $q$ $\leq$ $q_p$ we observe  
that the MC theory fairly accounts for the OF-AIMD results, with an 
accuracy comparable to what has already been achieved in 
other liquid metals \cite{Mont}. Consequently, the present 
results show the ability of the MC theory to describe the single particle 
dynamics (and presumably the collective dynamics too) in liquid systems 
encompassing a range of bonding and structure as that displayed by the 
compressed l-Si .

\section{Conclusions.}

Several static and dynamic properties of l-Si at four high-pressure 
thermodynamic states have been investigated using orbital free 
{\em ab initio} molecular dynamics combined with a first-principles local 
pseudopotential.

The study was motivated by experimental findings \cite{FunamoriTsuji} 
of significant structural changes in l-Si when the pressure is increased 
from 4 to 23 GPa. The obtained results for the static structure 
qualitatively follow those trends unveiled by the experiment, 
namely the increase of the intensity and the position of the 
S(q)'s main peak, along with a progressive vanishing of its shoulder. 
Other parameters such as the coordination number, isothermal 
compressibility, and the shape of the bond-angle distibution function  
provide further insight into the changes. Overall, 
apart from a contraction with 
increasing pressure, the static structures of l-Si at the TP, and at 4 
and 8 Gpa are very similar. Above 8 GPa the system transforms to a 
denser more close packed structure typical of a liquid simple metal, 
with most change taking place between 8 and 14 GPa.

The structural changes are also reflected in several dynamical 
properties. The calculated dynamic structure 
factors, $S(q, \omega)$, show collective density excitations 
over similar wavelength ranges, namely up to $q$ $\approx$ (3/5)$q_p$, 
as those found for simple liquid 
metals at their TP. These density excitations are sound waves whose 
velocity increases with pressure, most steeply between 8 and 14 GPa. 
The dispersion relations of the excitations divide into two groups,
one for l-Si at its TP, 4 GPa and 8 GPa and another group for l-Si 
at 14 and 23 GPa. 

The transverse current correlation also show evidence of the structural 
changes. Below 4 GPa, its frequency spectra lack inelastic peaks indicating
the absence of shear waves, but at 8 GPa clear inelastic peaks
are already evident. 

The calculated self-diffusion and shear viscosity transport coefficients
are also affected by the structural changes occurring between 8 and 14 
GPa. These transport coefficients cannot be compared with experiment, 
but confidence in the calculated values is given by the good agreement
with experimental values and/or other {\it ab-initio} results for l-Si
and its TP. 

Finally, we remark that the present results for the static and dynamic 
properties of compressed l-Si underscore the capability of the OF-AIMD 
method to tackle liquid systems encompassing a range of bonding and structure 
which evolves from mild remnants of covalent bonding to a metallic one. 
Moreover, further improvements in the 
present {\it ab initio} method are still possible and they 
necessarily will be focused on 
developing more accurate electron kinetic energy functionals and 
local ionic pseudopotentials.

\section*{Acknowledgements}

This work has been supported by the 
MEC of Spain (MAT2005-03415) and the 
NSERC of Canada. DJG acknowledges additional financial support from  
the Physics Dept. of Queen's University were part of this work was 
carried out.

\end{document}